\documentclass{article}
\usepackage[final, nonatbib]{nips_2017}

\usepackage[utf8]{inputenc} 
\usepackage[T1]{fontenc}    
\usepackage{hyperref}       
\usepackage{url}            
\usepackage{booktabs}       
\usepackage{amsfonts}       
\usepackage{nicefrac}       
\usepackage{microtype}      
\usepackage{graphicx}
\usepackage{ctable}

\usepackage{mathtools}
\usepackage{kotex}
\usepackage{times}
\usepackage{subfigure} 
\usepackage{titlesec}       
\usepackage{amsmath}
\usepackage{amssymb}
\usepackage{algorithm}
\usepackage[noend]{algpseudocode}
\usepackage{mathptmx,siunitx}

\makeatletter
\def\algbackskip{\hskip-\ALG@thistlm}
\makeatother

\title{Homomorphic Parameter Compression for Distributed Deep Learning Training}

\author{
    Jaehee Jang, Byunggook Na, Sungroh Yoon \\
    Department of Electrical Engineering \\
    Seoul National University \\
    \textit{sryoon@snu.ac.kr}
}

\begin{document}

\maketitle

\begin{abstract}
Distributed training of deep neural networks has received significant research interest, and its major approaches include implementations on multiple GPUs and clusters. Parallelization can dramatically improve the efficiency of training deep and complicated models with large-scale data.
A fundamental barrier against the speedup of DNN training, however, is the trade-off between computation and communication time. In other words, increasing the number of worker nodes decreases the time consumed in computation while simultaneously increasing communication overhead under constrained network bandwidth, especially in commodity hardware environments. 
To alleviate this trade-off, we suggest the idea of homomorphic parameter compression, which compresses parameters with the least expense and trains the DNN with the compressed representation. Although the specific method is yet to be discovered, we demonstrate that there is a high probability that the homomorphism can reduce the communication overhead, thanks to little compression and decompression times. We also provide theoretical speedup of homomorphic compression.\end{abstract}

\section{Introduction}

Deep learning (DL) derives structured information from raw data using deep neural networks (DNN). DL finds hierarchical representations of data through several non-linear layers of a DNN. When the problem to be solved by using DL is challenging, we need to grasp complicated representations from the data. With the use of DNNs to solve an increasing number of high-abstraction problems in various fields, the size of training models and the computational load to train the models have continued to grow. Under current software and hardware constraints, DNN training demands a massive amount of processing time~\cite{keuper2016distributed}, naturally leading to the need for distributed deep learning naturally uprose~\cite{bekkerman2011scaling,dean2012large,recht2011hogwild,kim2016deepspark,tensorflow2015-whitepaper}. Distributed DL divides the workload (training data or model) to different machines and aims for faster learning while maintaining the original performance of the model.

\begin{figure}[t]
    \centering
    \includegraphics[width=.9\textwidth]{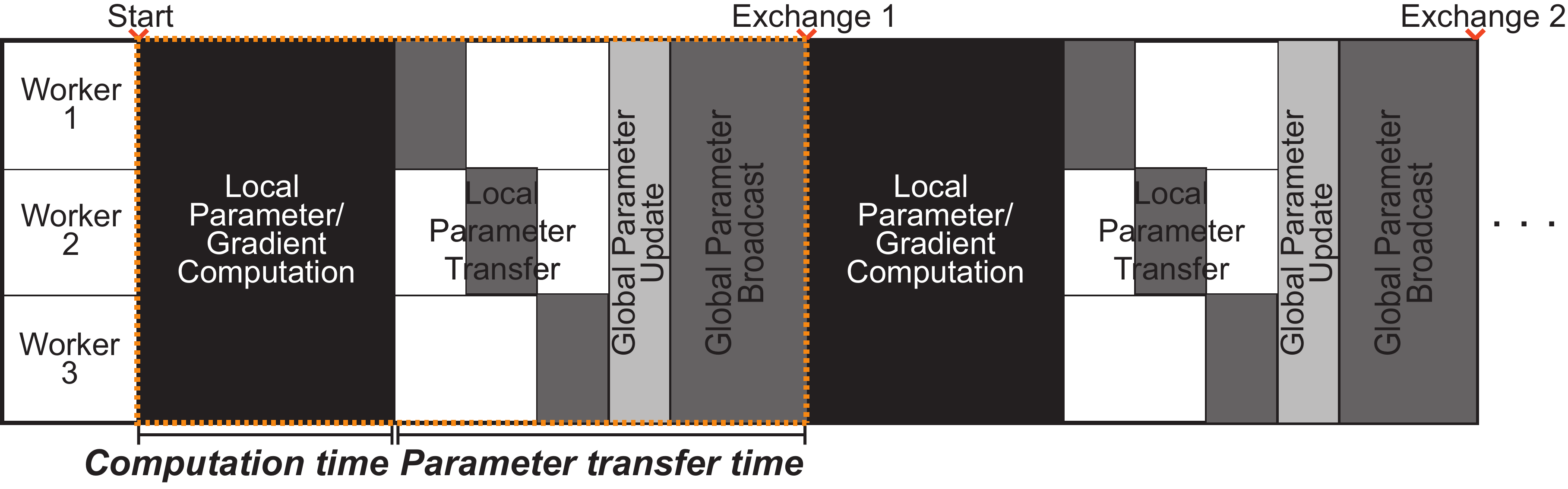}
    \label{fig:distdl}
    \caption{Simplified process demonstration of distributed DL using synchronous SGD}
\end{figure}

DNN iteratively optimizes weight parameters based on gradients computed from feedforward/backpropagation, which is highly sequential. Hence the implementation of distributed DNN training requires specific design principles and strategies as they have been suggested for years~\cite{xing2016strategies}. To give a brief illustration on how distributed DL is implementated in general, let's take the synchronous SGD update scenario as an example (Fig.~\ref{fig:distdl}). Synchronous SGD trains by iterating a set of processes to update global parameters, described by a dotted box in Fig.~\ref{fig:distdl}. The set of global parameter update consists of the following steps. First, all the worker nodes train until the designated number of iterations (\textit{Local Parameter/Gradient Computation} in Fig.~\ref{fig:distdl}). Then, all the worker nodes respectively push their local parameters to the parameter server (\textit{Local Parameter Transfer} in Fig.~\ref{fig:distdl}). Lastly, the parameter server decides global parameters by aggregating all the pushed local parameters(\textit{Global Parameter Update} in Fig.~\ref{fig:distdl}), and pulls them to the worker nodes(\textit{Global Parameter Broadcast} in Fig.~\ref{fig:distdl}).

 The worker nodes participating in the training frequently exchange their training status with other nodes so that the model can reflect all the divided workloads. However, DNN models are large and so the communication load. Hence they cause bottlenecks on transmission because of the contrained communication bandwidth. Especially under commodity hardware environments, the weight-transfer time overwhelms even the computing time. Along with the transmission time, the time technically required for communication, that is, the time to prepare and sustain communication, is also included as communication overhead. The communication overhead is one of the main factors that increase the parallel training time. In order to alleviate the communication overhead, attempts have been made to reduce the model size before communication~\cite{iandola2016squeezenet,spring2016scalable,elgohary2016compressed}.

The goal of our study is to demonstrate \textit{homomorphic parameter compression}, which is a novel concept of compressed deep learning. As the term \textit{homomorphic} suggests, it is a compression method that reduces the size of parameters and allows key operations of DL to be executed without decompression. Since parameters are transferred numerous times during distributed training settings, this method can remarkably reduce the time consumed in communication, which is the main rate-limiting step of distributed training. Furthermore, homomorphism prevents the generation of additional overhead by repetitive compression and decompression. The main contributions of this paper include the followings: 1) To our knowledge, this is the first attempt to demonstrate homomorphic parameter compression. 2) We theoretically characterize the possible factors in the parameter compression, e.g., the compression ratio, and provide thorough simulative analyses. 3) We provide the theoretical reduction in training time of the homomorphically compressed distributed training in function of the number of participating worker nodes for different values of the compression ratio.

\section{Literature Survey on Compressed Deep Learning}
\label{sec:compressed-dl}

Numerous studies have suggested \textit{compression} in deep learning~\cite{han2015deep,courbariaux2015binaryconnect,courbariaux2016binarized,seide20141}. Existing compression methods aim at fitting very-large-scale models into a mobile device or single FPGA chip, at alleviating the high communication overhead due to distributed training, and at improving computational performance as well as storage and power efficiency.

\textbf{Post-training compression for inference.}\hspace{0.2cm} 
A series of studies reduced the storage and energy required to run inference on large DL models and deploy them on embedded systems or mobile devices. Deep compression~\cite{han2015deep} used pruning, trained quantization, and Huffman encoding on weights and demonstrated a high compression ratio to fit in on-chip memory. CNNpack~\cite{NIPS2016_6390} demonstrated convolutional neural network (CNN) compression in the discrete cosine transform (DCT) frequency domain. These methods can effectively reduce the size of networks while retaining pre-trained information. On the other hand, since they are designed for compression after training is completed, the time consumed in compression is a minor issue.

\begin{figure}[t]
    \centering
    \includegraphics[width=.9\textwidth]{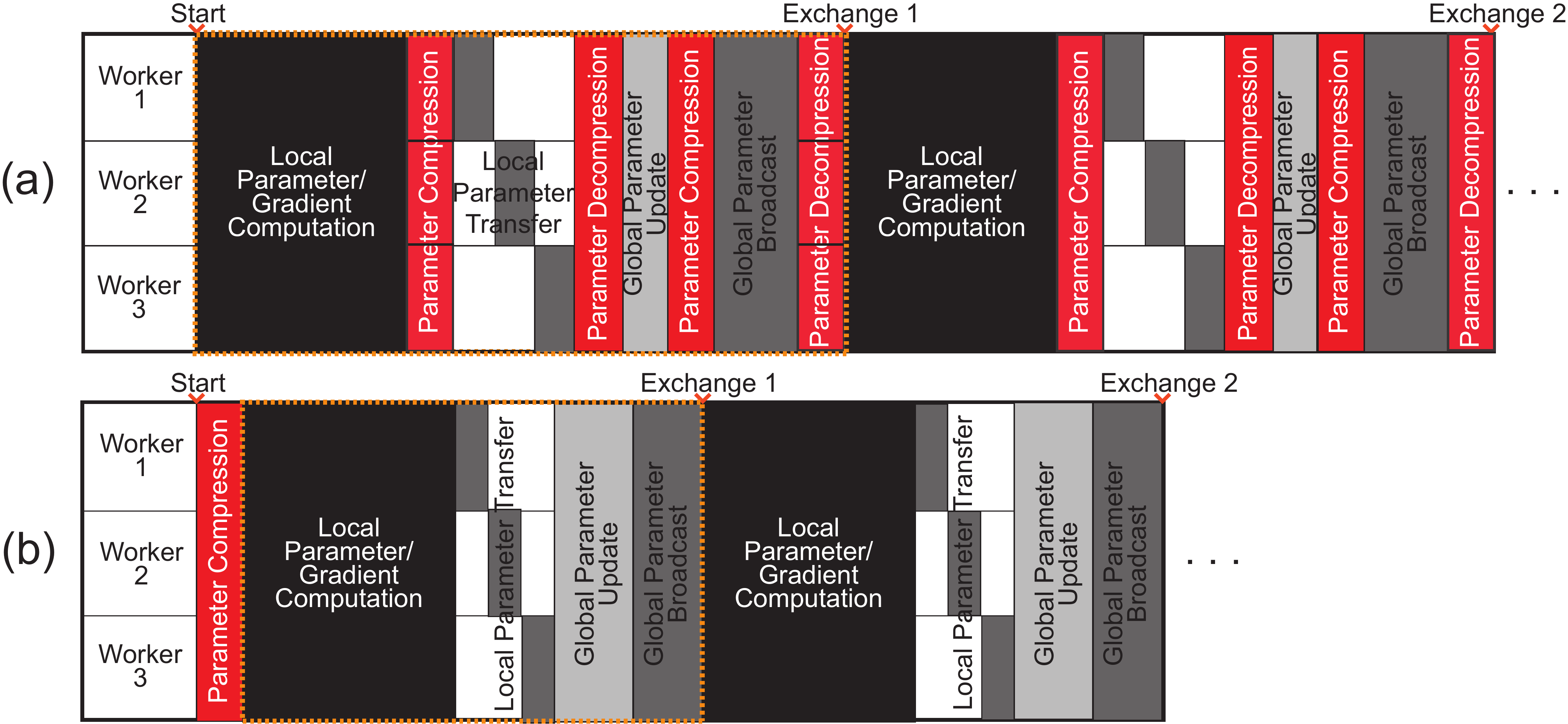}
    \label{fig:methods}
    \caption{Simplified process demonstration of mid-training compression methods, based on synchronous SGD: (a) repetitive (de)compression, and (b) one-time compression. Note that we considered \textit{parameter transfer time} as the time in which a worker node is not training but waiting for new gradient update. (Best viewed in color)}
\end{figure}

\textbf{In-training compression for refficient deep learning.} \hspace{0.2cm}
 Employing compression in training enables efficient computation and communication under limited resources. Especially when transferring parameters in distributed DL, the constrained network bandwidth may consume a large amount of time in communication and slow down the entire training process. The following approaches proposed compression for both training and inference in order to improve computational performance as well as energy and storage efficiency. We classified the approaches into two types: repetitive (de)compression and one-time compression. The training process of the methods is illustrated in Fig.~\ref{fig:methods}.

\textbf{Repetitive (de)compression}\hspace{0.2cm} 
Some methods encode weights (or gradients) for every iteration We call such methods \textit{repetitive (de)compression} methods. Weight binarization methods~\cite{courbariaux2015binaryconnect,courbariaux2016binarized} binarize weights in order to train from low-power devices or specialized hardware. The binarized weights are used only during the forward and back propagations but not during parameter update. The authors of \cite{seide20141, alistarh2016qsgd} used gradient binarization with distributed DL training in order to reduce the communication overhead. They encoded gradients during global parameter update, and worker nodes have to decode the gradients to update their local parameters. FreshNet~\cite{Chen:2016:CCN:2939672.2939839} combined hashing~\cite{Chen:2015:CNN:3045118.3045361} with DCT to compress CNN models and train in the frequency domain. Good compression performance and robustness in model accuracy were demonstrated . However, continuous compressions and decompressions involve high risks for additional compression overhead, as shown in Fig.~\ref{fig:methods}(b). Although good compression performance was demonstrated, more careful considerations are needed to utilize the aforementioned methods in distributed training, as is done with QSGD~\cite{alistarh2016qsgd} by double buffering. A quantitative analysis of compression overhead will be presented in Section~\ref{sec:experimental}.

\textbf{One-time compression}\hspace{0.2cm} 
If DL models are compressed only once, the compression time will not significantly affect the overall training time. Compressed linear algebra (CLA)~\cite{elgohary2016compressed} exploits lightweight database compression techniques to compress matrices and perform computations in the compressed representation. Despite the compression ratio and the operation performance being close to that of uncompressed operations, it is difficult to be applied directly to distributed DL training because more nonlinear operations are required in DL training. Especially, operations that are frequently used in DL training, such as normalization and pooling, are not yet conducted in the compressed representation~\footnote{The confirmed version of September 2017 can be found at https://github.com/apache/systemml}.

\section{Algorithmic Design}
\label{sec:algorithmic-design}
\begin{table}[t]
\centering
\caption{Examples of Gzip compression}
\label{tb:comp}
\addtolength{\tabcolsep}{-0.3pt}
\renewcommand{\arraystretch}{1}
\small
\begin{tabular}{
  @{}l
  S[table-format=1]
  S[table-format=1.3]
  S[table-format=5.3]
  S[table-format=4.3]
  @{}
}
\toprule
{} &{ \textbf{Size}} & { \textbf{Compression}} & { \textbf{Compression}} & { \textbf{Decompression}} \\
{} & {} & { \textbf{ratio}} & { \textbf{time}} & { \textbf{time}} \\
\midrule
\centering
\textbf{AlexNet Caffemodel} & 233MB & 1.079 &  8.079s &  1.898s \\
\textbf{ILSVR2012 Train Data} & 240GB & 1.269 &  10532.177s &  3886.498s \\
\textbf{CIFAR-100 Train Data} & 147MB & 1.097 &  660.876s &  2.242s \\
\bottomrule
\end{tabular}
\vspace*{-0.5cm}
\end{table}
Communication overhead is one of the major drawbacks of distributed DL, and compressing the parameters can reduce the communication workload. On the other hand, if the parameters are compressed and decompressed at every update, there are high risks for additional \textit{compression overhead}. As indicated in Table~\ref{tb:comp}, compressing and decompressing parameters can take considerable time. Therefore, we need a compression approach that does not increase the computing time significantly while reducing the size of parameters properly.

\begin{figure}[t]
    \centering
    \includegraphics[width=.6\textwidth]{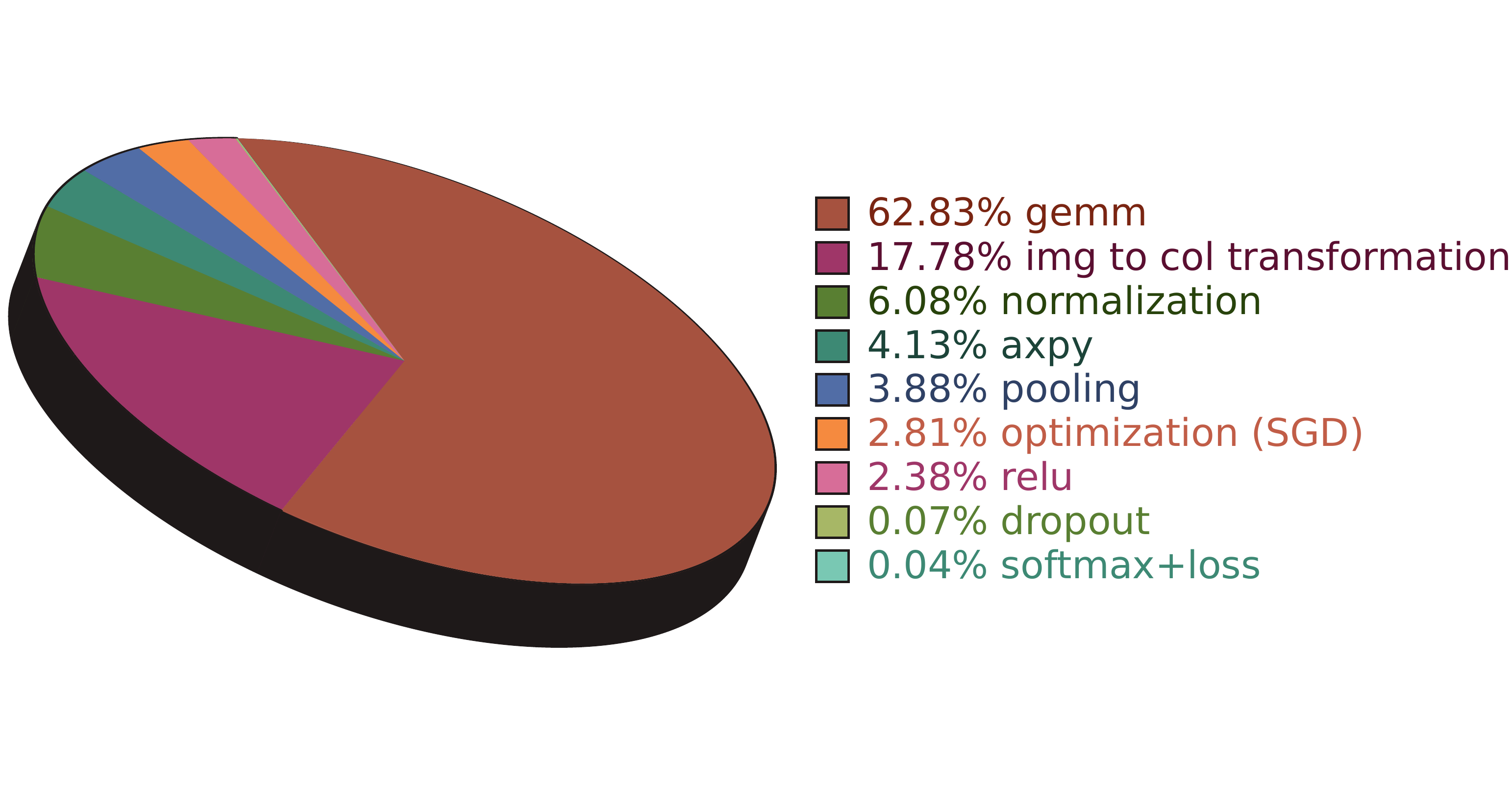}
    \caption{GPU kernel analysis of AlexNet training (Caffe)}
    \label{fig:operation}
\end{figure}

 We propose \textit{homomorphic parameter compression}, inspired from homomorphic encryption~\cite{rivest1978data}. Homomorphism suggests an algebraic system that is encoded from another algebraic system and performs operations equivalent to those of the encoded system. Our goal is to propose a compression method that can be trained without decompression. By referring to the early formulation of homomorphic encryption~\cite{rivest1978data}, the definition of homomorphic compression is as follows. Suppose we have a system $S=<W;f_{1},f_{2},...>$ that consists of a set of parameters $W$ and operations $f_i$s concerned with training. The possible $f_i$s may vary depending on the model structure. As Fig.~\ref{fig:operation} shows, linear operations take the majority of the operations and nonlinear operations such as pooling and relu are included as well. We propose finding the encoding function $\phi : S \rightarrow S'$, where $S'=<W';f_{0}',f_{2}',...>$, where  is the compressed system.

\begin{enumerate}
    \item \textbf{An encoded version of a weight $w_{i}'=\phi(w_{i})$ should be smaller than original weight $w_{i}$.}
    \item \textbf{$\phi$ should be easy to compute.}
    Conversion by $\phi$ should not take too much time. We point out how compression overhead can slow down the total training time in Fig.~\ref{fig:cpd}. We can continue training even without decompression if the compression time is long enough to affect the total training time, as it is difficult to expect temporal gain through homomorphic compression in such a case.
    \item \textbf{The operations $f_{i}'$ should be efficiently computable.} 
    When training DL, varied operation functions ($f_{i}$s) are required, such as matrix multiplication and activation functions, as shown in Fig. 2 in Supplement. If we encode the functions to equivalent $f_{i}'$ operations, the computational efficiency of $f_{i}'$ operations is also required to be high.
\end{enumerate}

\section{Experimental Study}
\label{sec:experimental}
\subsection{Experimental Settings}

\textbf{Notations \& formulated assumptions}
We assumed of a distributed training environment where there are M worker nodes. It parallelizes a single-node training with minibatch size B on target dataset of size D. The single node consumes C of time when computing a minibatch, and the total size of weight parameters is measured as W.

\textbf {Optimization Scheme}
When conducting distributed training, we can define various optimization schemes according to when local parameters have been updated (worker nodes have been trained) with global parameters (parameters that all worker nodes share). Communication overhead is inevitable regardless of the strategy we shall choose. In order to emphasize the effect of communication overhead with different numbers of worker nodes, we assumed that we optimize the DNN training by synchronous stochastic gradient descent (synchronous SGD).

Synchronous SGD trains by iterating a set of processes to update global parameters, as described in Fig.~\ref{fig:distdl}. In this paper, we defined the time required in the local parameter/gradient computation step until the designated number of iterations $i$ as \textit{computation time, $T_{cmt}$}, and the time required in the remaining steps as \textit{parameter transfer time, $T_{tnf}$} which adds up to the time required for one set of global parameter update, $T_{update} = T_{cmt} + T_{tnf}$.

\textbf {Minibatch size, $B \rightarrow b$} 
We assumed data-parallelized training, which divides the training dataset $D$ into $M$ worker nodes. As the dataset is divided, the ratio of the original minibatch size to training data size becomes larger. If the batch size is too large, the training may be delayed~\cite{bengio2012practical}. Hence it is assumed that the minibatch size is reduced by the reduction amount of the training set. Therefore, \(b = \frac{B}{M}\).

\textbf {Minibatch computation time, $C \rightarrow c$}
As the minibatch size decreases by $\frac{1}{M}$, the time consumed for one iteration is expected to decrease by the same ratio. Therefore, \(c = \frac{C}{M}\)

\textbf {Computation time per update, $T_{cmt}$}
Since the data and minibatch size are decreased by $\frac{1}{M}$, training iterations are required to be tje same as single-node training in order to train the same number of epochs. Hence, \(T_{cmt} = i \cdot c = \frac{i \cdot C}{M}\).

\textbf {Parameter transfer time,  $T_{tnf}$}
In synchronous SGD, every local parameter of the worker node is collected when updating a global parameter. If the number of participating worker nodes increases, the number of parameters to be exchanged linearly increases. That is, if we define the size of the weight parameter as $W$ and train with $M$ worker nodes, the number of parameters required to communicate is $W \cdot M$. 
Letting the transmission rate of the cluster be $\chi$, the parameter transfer time is denoted as \(T_{tnf} = \frac{W \cdot M}{\chi}\).

From the assumptions stated in Section~\ref{sec:algorithmic-design}, we simulated distributed training for CNN and RNN models. We trained the ImageNet dataset~\cite{deng2009imagenet} using AlexNet~\cite{krizhevsky2012imagenet} to simulate distributed CNN training. For simulative distributed RNN training, we conducted STREET model~\cite{Smith2016} training on the French Street Name Signs (FSNS) dataset~\footnote{TensorFlow implementation at https://github.com/tensorflow/models/tree/master/street}. We used Caffe~\cite{jia2014caffe} for the CNN model and TensorFlow~\cite{tensorflow2015-whitepaper} for the RNN model as computing engines. Synchronous SGD iterates the same processes of global parameter update, where the worker nodes train up to the designated iterations and communicate parameters to apply the global training trend. Hence, Figs.~\ref{fig:dist},~\ref{fig:cpd},~\ref{fig:bound}, and Fig.~\ref{fig:hpc}(b), (c) represent only one set of global updates for the total training trend in terms of time.

The hardware we used in the simulation is a commodity cluster. We used a homogeneous cluster consisting of 25 identical machines connected via Gigabit Ethernet. Each worker node has an Intel Core i7-4790 processor with 16GB of main memory and an NVIDIA GTX970 GPU.

\subsection{Simulated Analysis of Communication Overhead Effect and Na\"ive Parameter Compression}
\label{sec:simulated-analysi}
\begin{figure}[t]
    \centering
    \includegraphics[width=.95\textwidth]{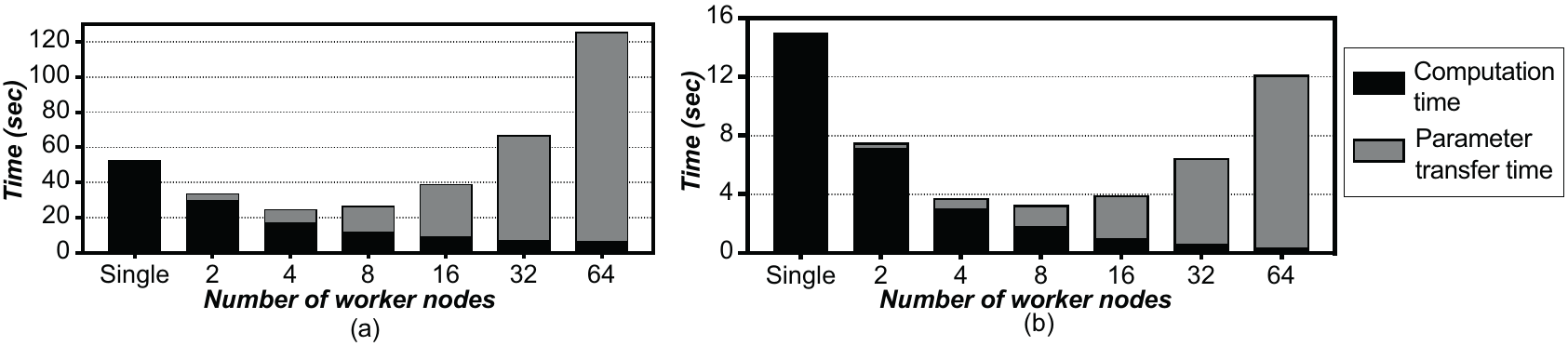}
    \caption{Simulated analysis of $T_{cmt}$ vs. $T_{tnf}$ on vanilla synchronous SGD training. We conducted experiments for one set of global parameter update. (a) 200-iter AlexNet training (Caffe). (b) 20-iter STREET training (TensorFlow).}
    \label{fig:dist}
\end{figure}
Fig.~\ref{fig:dist} shows the simulated global parameter update of vanilla synchronous SGD. As the number of nodes increases, the computation time decreases but the parameter transfer time increases. At some points, parameter transfer takes more time than minibatch computation, which demonstrates the serious inefficiency of resource utility due to communication overhead in distributed training. If we keep increasing the number of nodes, the parameter transfer time even exceeds the single minibatch computation time, becoming a hindrance rather than contributing to speedup.

\begin{figure}[t]
    \centering
    \includegraphics[width=.95\textwidth]{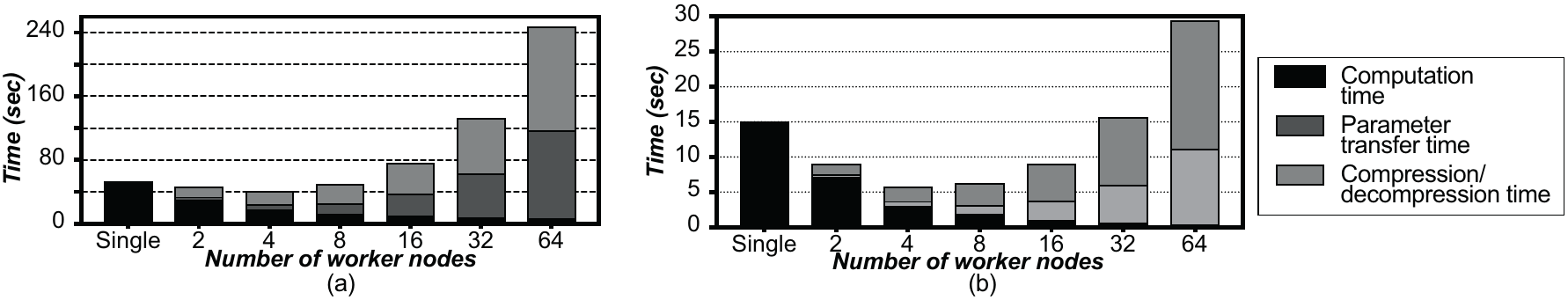}
    \caption{Simulated analysis of $T_{cmt}$ vs. $T_{tnf}$ on vanilla synchronous SGD training with a common compression method. We conducted experiments for one set of global parameter update. We assumed Gzip compression/decompresion at every parameter update. (a) 200-iter AlexNet training (Caffe). (b) 20-iter STREET training (TensorFlow).}
    \label{fig:cpd}
\end{figure}
If we can compress weight parameters by $\rho$x, the parameter transfer requirement will be reduced by $\frac{1}{r}$. This will also reduce the time required for parameter transfer. However, the time consumed in compressing parameters is a problem. Suppose after a worker node finished its batch, it compresses the trained parameter using Gzip. Then, the parameter server aggregates the compressed parameters and since Gzip-compressed data are not computable, the parameter server has to decompress the parameters to aggregate and average. After global parameters are set, the parameter server compresses the parameters again and sends them back to the worker nodes. When a worker node receives the compressed parameters, it has to decompress them again to keep training.

If compressed training is conducted as explained above, there is a high possibility of another overhead called compression overhead. We simulated this in Fig.~\ref{fig:dist}, and the results are shown in Fig.~\ref{fig:cpd}. If we use Gzip compression, the parameter transfer time is decreased because of the reduced size of parameters. On the other hand, the compression and decompression time can significantly exceed the original training time, as shown in Fig.~\ref{fig:cpd}.
\subsection{The Expected Gain of Homomorphic Compression}

From the simulation results in Section~\ref{sec:experimental}, we can learn two things. First, compressing the parameter size can reduce the communication workload, and second, we need a compression approach that takes compression overhead into account. Therefore, we propose homomorphic parameter compression here.

\begin{figure}
    \centering
    \includegraphics[width=.95\textwidth]{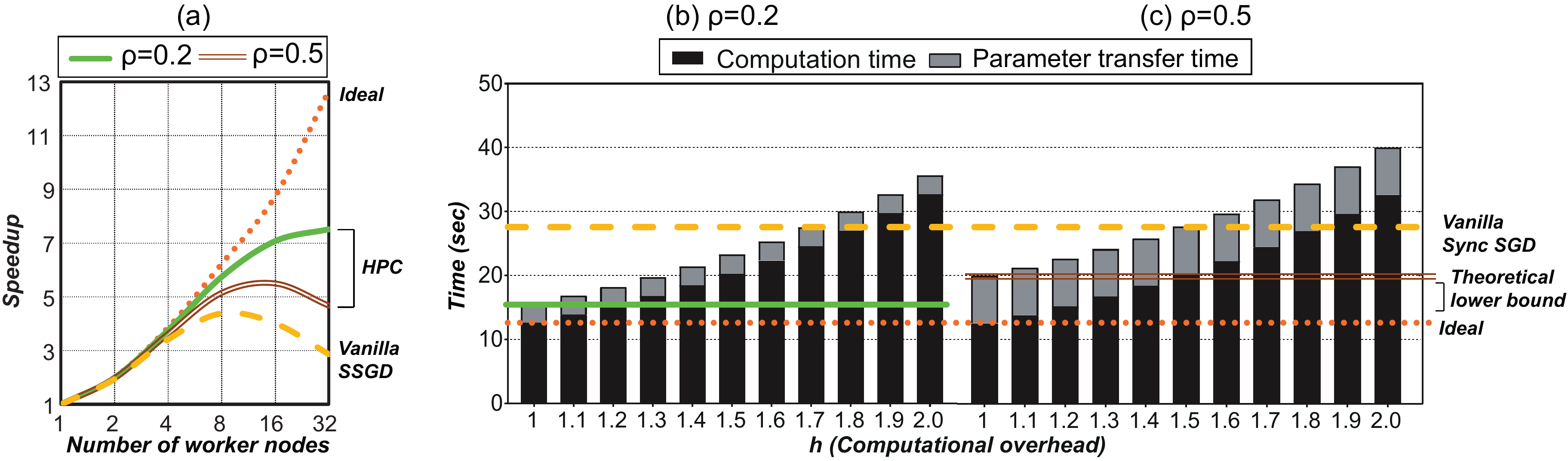}
    \caption{(a) Theoretical speedup as a function of the number of worker nodes for different compression ratios when training the ImageNet with AlexNet. (b)\&(c) Simulated analysis of the proposed method for 200-iter AlexNet training on 16 worker nodes, when $h=1.0, 1.1, ... 2.0$ and (b) $\rho=0.2$. and (c) $\rho=0.5$. We illustrated the expected $T_{cmt}$ and $T_{tnf}$ for one set of global parameter update. Compression and decompression are not shown in these graphs because they are only performed once throughout the training.(Best viewed in color)}
    \label{fig:bound}
\end{figure}

Fig.~\ref{fig:bound}(a) shows the theoretical speedup of homomorphically compressed distributed training based on the simulative analysis conducted in this section. It is represented as a function of the number of participating worker nodes for different compression ratios. The orange dotted curve in Fig.~\ref{fig:bound}(a) shows the ideal distributed training case, where there is no communication at all so we can achieve linear speedup. The yellow dashed curve is the speedup of vanilla SGD. The ideal speedup in homomorphic compression occurs when there is no overhead due to the increased operation time. Hence, the green solid and red double solid curves are the theoretical upper bounds in speedup when the compression ratio is 0.2 and 0.5 respectively. 

\begin{figure}[t]
    \centering
    \includegraphics[width=.95\textwidth]{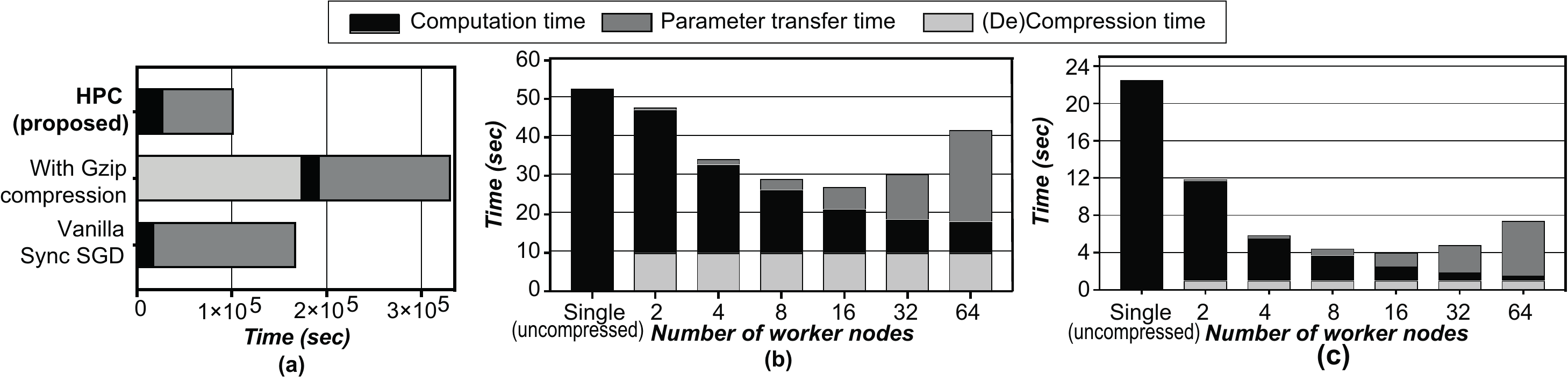}
    \caption{Simulated analysis of the proposed method. (a) Comparison of total Alexnet training time among the proposed method, synchronous SGD with Gzip compression, and vanilla synchronous SGD. (b)\&(c) $T_{cmt}$ vs. $T_{tnf}$ on synchronous SGD training with proposing method. We illustrated the expected result of one set of global parameter update. Compression and decompression shown in the graphs are performed only once throughout the training. (b) 200-iter AlexNet training (Caffe) when $h=1.3$, $\rho=0.2$. (c) 20-iter STREET training (TensorFlow) when $h=1.5$, $\rho=0.5$.}
    \label{fig:hpc}
\end{figure}

In actual homomorphic compression, the encoded operations are likely to require more time than the original operations. The simulated results based on this assumption are shown in Fig.~\ref{fig:hpc}. Note that the compression time in Fig.~\ref{fig:hpc} is not included in every parameter update but only in the early stage. We assumed that computing in the compressed representation may take more time than the original computations. However, if we can manipulate the operation overhead and compression ratio, fast and large-scale DL training is attainable, as shown in Fig.~\ref{fig:hpc}.

\section{Discussion and Future Work}
We analyzed the effect of homomorphic compression on distributed training in Section~\ref{sec:experimental}. In this section, we discuss the parameters required when designing a homomorphic compression method. Below, we present an in-depth analysis on the computational efficiency for $f_{i}'$'s based on the assumptions made in Section~\ref{sec:algorithmic-design} and Section~\ref{sec:experimental}: 

Let $\rho$ be a compression ratio, \(\rho = \frac{\phi(w_{i})}{w_i}\) (where \(0<\rho<1\)), and \(T_{f_{i}})\) and \(T_{f_{i}'}\) be the time consumed in performing \(f_{i}\) and \(f_{i}'\) respectively. And we define \textit{operation overhead} $h$ as \(h = \frac{T_{f_{i}'}}{T_{f_{i}}}\). Then, the compressed minibatch computation time $T_{cmt}'$ and parameter transfer time $T_{tnf}'$ are expressed as $ C \cdot T_{cmt}' = \frac{C}{M} \cdot h $, $ T_{tnf}' = \frac{MW}{BW} \cdot \rho $, respectively.

By setting the upper bound of the total training time as \(\frac{C}{M} \cdot r\) (where \(1\leq r \leq M\)), the relationship between h and $\rho$ can be obtained as expressed by Eq.~\ref{eq:h}. Therefore, we can achieve the desired speedup if we can fit $h$ and $\rho$ under Eq.~\ref{eq:h}.

\begin{align*} \label{eq:h}
    \frac{C}{M} \cdot r \geq \frac{C}{M} \cdot h + \frac{MW}{BW} \cdot \rho \\
    \therefore h \leq -\frac{M^{2}W}{CBW} \cdot \rho + r 
\end{align*}

Fig.~\ref{fig:bound}(b) expected tendency of training time with respect to $h$ and $\rho$. The most ideal training time for $M$ worker nodes to learn a model that takes batch computation time $C$ is $\frac{C}{M}$ (orange dotted line in Fig.~\ref{fig:bound}(b)). However, even when we train in parallel, the learning status among nodes is interchanged. Therefore, the realistic training time of synchronous SGD, considering communication time, is the same as the red line in Fig.~\ref{fig:bound}(b).

In addition to the upper bound the computation time for $f_{i}'$, $S'$ and $\phi$ are needed to be designed in consideration of frequently used operations. Fig. 8 shows the GPU profiling result of AlexNet training with Caffe, and it suggests that operations such as gemm  take most of the computation time. It is expected that, if we significantly reduce the time required for computing gemm, the operation overhead effect will be much weaker. Our future work is to propose a detailed homomorphic compression method.


\end{document}